\documentclass[10pt,aps,prl,amsmath,amssymb,nobibnotes,longbibliography,superscriptaddress,twocolumn]{revtex4-2}

\usepackage{ulem}
\usepackage[dvipsnames]{xcolor}
\usepackage{graphicx}
\usepackage{dcolumn}
\usepackage{bm}
\usepackage{amssymb}
\usepackage{amsmath}
\usepackage[unicode=true, colorlinks=true, citecolor={blue!80!black}, urlcolor={blue!50!black}, linkcolor = {blue!80!black}]{hyperref}
\usepackage{physics}
\usepackage{siunitx}
\usepackage{float}
\usepackage[english]{babel}

\begin{document}
\preprint{APS/123-QED}

\title{Ultrastrong coupling with a gate-tunable transmon}
\title{Ultrastrong Coupling and Coherent Dynamics in a Gate‑Tunable Transmon Qubit}

\newcommand{\affFMC}{\affiliation{Departamento de F\'{i}sica de la Materia Condensada, Universidad Aut\'{o}noma de Madrid, 28049 Madrid, Spain}}

\newcommand{\affFTMC}{\affiliation{Departamento de F\'{i}sica Te\'{o}rica de la Materia Condensada, Universidad Aut\'{o}noma de Madrid, 28049 Madrid, Spain}}

\newcommand{\affIFIMINC}{\affiliation{Condensed Matter Physics Center (IFIMAC) and Instituto Nicolás Cabrera (INC), Universidad Aut\'{o}noma de Madrid, 28049 Madrid, Spain}}

\newcommand{\affICMM}{\affiliation{Instituto de Ciencia de Materiales de Madrid (ICMM), Consejo Superior de Investigaciones Científicas (CSIC), Sor Juana Inés de la Cruz 3, 28049 Madrid, Spain}}

\newcommand{\affBOHR}{\affiliation{Center for Quantum Devices, Niels Bohr Institute, University of Copenhagen, 2100 Copenhagen, Denmark}}

\newcommand{\affQTech}{\affiliation{QuTech and Kavli Institute of Nanoscience, Delft University of Technology, 2600 GA Delft, Netherlands}}

\newcommand{\affINTA}{\affiliation{Centro de Astrobiología (CSIC - INTA), Torrej\'{o}n de Ardoz, 28850 Madrid, Spain}}

\newcommand{\affIMDEA}{\affiliation{IMDEA Nanociencia, Cantoblanco, 28049 Madrid, Spain}}

\newcommand{\affLAB}{\affiliation{Laboratorio de Transporte Cu\'{a}ntico, Unidad Asociada UAM/ICMM-CSIC, Madrid, Spain}}

\author{I. Casal Iglesias}
\affFMC \affIFIMINC

\author{F. J. Matute-Cañadas}
\thanks{These authors contributed equally to this work}
\affIFIMINC \affFTMC

\author{G. O. Steffensen}
\thanks{These authors contributed equally to this work}
\affICMM \affLAB

\author{A. Ibabe}
\affFMC \affIFIMINC

\author{L. Splitthoff}
\affQTech

\author{T. Kanne}
\affBOHR

\author{J. Nyg\aa rd}
\affBOHR

\author{V. Rollano}
\affINTA

\author{D. Granados}
\affIMDEA

\author{A. Gomez}
\affINTA

\author{R. Aguado}
\affICMM \affLAB

\author{A. Levy Yeyati}
\affIFIMINC \affFTMC \affLAB  

\author{E. J. H. Lee}
\email{eduardo.lee@uam.es}
\affFMC \affIFIMINC \affLAB

\date{\today}

\begin{abstract}

Ultrastrong light–matter coupling (USC) gives access to exotic quantum phenomena and promises faster quantum gates, 
yet coherent time-domain control in this regime remains largely unexplored. Here, we realize USC in a hybrid system consisting of an InAs nanowire-based gatemon qubit coupled to a superconducting resonator. Spectroscopy reveals an avoided crossing that cannot be captured by the Jaynes-Cummings (JC) model, as well as photon-number-dependent transitions whose energies deviate markedly from the JC ladder expected in the strong coupling regime. Beyond demonstrating USC, we achieve time-resolved coherent control of the qubit and measure coherence times comparable to gatemons operating outside the USC regime. These results establish that hybrid semiconductor-superconductor qubits can retain coherent control in USC and provide a platform for exploring quantum dynamics and device concepts in this regime.



\end{abstract}

\maketitle

\textit{Introduction.-} Since the demonstration of strong coupling (SC) in superconducting circuits~\cite{Wallraff2004}, circuit quantum electrodynamics (cQED) has emerged as a powerful platform for studying artificial atoms coupled to individual electromagnetic modes at the single-photon level~\cite{Schuster2007,Bishop2008,blais_circuit_2021}.  By further increasing the strength of the light-matter interaction, superconducting circuits have entered the ultrastrong coupling (USC) regime~\cite{Devoret2007,FornDiaz10,Niemczyk2010,Bosman2017PRB,Bosman2017npj,FornDiaz16,Yoshihara16,wang_probing_2023}. Here, the interaction strength becomes a significant fraction of the bare system frequencies, with $g/\omega_r > 0.1$ ($g$ being the coupling strength and $\omega_r$ the resonator angular frequency), leading to the breakdown of the rotating-wave approximation (RWA) and the emergence of qualitatively new physical phenomena~\cite{FornDiaz19,Kockum2019}. However, despite the potential for faster quantum gates~\cite{romero_ultrafast_2012,wang_ultrafast_2017}, experimental realizations of the USC regime remain limited and time-domain coherence measurements, central to quantum information applications~\cite{peropadre_switchable_2010,ashhab_qubit-oscillator_2010,garziano_multiphoton_2015,Stassi25}, have remained elusive.

\begin{figure}[t!]
\includegraphics[width=86mm]{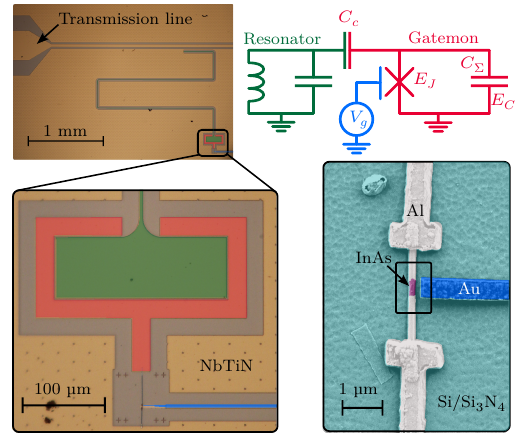}
\caption{\label{fig:sample_geometry} False-color optical and SEM micrographs of the gatemon–resonator device. Zoomed views highlight the superconducting island (red) and the coplanar waveguide resonator (green). The SEM image shows an InAs/Al nanowire junction with a side gate (dark blue), taken from a device similar to the one discussed in the main text; the junction length of the device studied here is approximately 200--250~nm. The circuit schematic defines the variables used in the model.}
\end{figure}

The above advances in USC have so far relied exclusively on aluminum oxide tunnel junctions. In parallel, cQED has increasingly incorporated semiconductors into superconducting circuits, giving rise to hybrid semiconductor–superconductor quantum devices~\cite{pita-vidal_novel_2025}. A prominent example is the gate-tunable transmon, or gatemon~\cite{larsen_semiconductor-nanowire-based_2015,de_lange_realization_2015}, in which the conventional tunnel junction is replaced by a semiconductor weak link, allowing electrostatic control of the qubit transition frequency. Moreover, the mesoscopic properties of such weak links lead to deviations from the standard Josephson potential with a cosine energy-phase relation~\cite{Furusaki1991,Beenakker1992,Bagwell1992,MartinRodero1994}, which modify the qubit level structure, for example 
by suppressing charge dispersion and reducing anharmonicity~\cite{kringhoj_anharmonicity_2018,kringhoj_suppressed_2020,bargerbos_observation_2020,aparicio_gate-tunable_2025,Fatemi25,feldstein-bofill_controlled_2026}.

In this work, we demonstrate ultrastrong light–matter coupling with a hybrid semiconductor–superconductor gatemon qubit. Spectroscopic measurements directly reveal a vacuum Rabi splitting consistent with the USC regime with $g/\omega_r$ values up to 0.2. In addition, we observe photon-number-dependent transitions whose energies clearly deviate from the Jaynes-Cummings (JC) ladder expected in the strong coupling regime. We study this feature quantitatively for the first time in a transmon architecture 
and observe that it arises due to the combination of USC and the multilevel qubit structure. 
Finally, time-resolved measurements demonstrate coherent qubit manipulation, with coherence times comparable to gatemons operated in the SC regime~\cite{larsen_semiconductor-nanowire-based_2015,casparis_gatemon_2016,danilenko_few-mode_2023,feldstein-bofill_gatemon_2024,sabonis_destructive_2020}. Our results showcase phenomena unique to the USC regime and demonstrate that, in gatemons, an accurate description must also include the mesoscopic properties of the weak link.






\textit{Sample description.-}
Our gatemon consists of a superconducting island (shown in red in Fig.~\ref{fig:sample_geometry}) 
connected to a superconducting reservoir through an InAs/Al nanowire-based Josephson junction. The qubit is capacitively coupled to a $\lambda$/4 coplanar waveguide resonator (green), which facilitates reaching the USC regime due to its highly confined electric field localized at its open end. Sample readout is performed through the inductive coupling of the resonator to a transmission line. A voltage, $V_g$, applied to a nearby side gate (dark blue) controls the critical current of the junction. 

\begin{figure}[b]
\includegraphics[width=86mm]{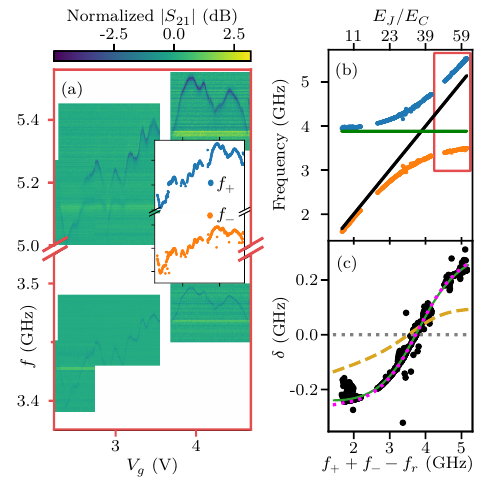}
\caption{\label{fig:demonstrating_usc} \textbf{(a)} Single-tone spectroscopy as a function of $V_g$. The colormap shows the normalized transmission magnitude, $|S_{21}|$ after background noise subtraction. The inset shows the extracted resonances $f_+$ (blue) and $f_-$ (orange). \textbf{(b)} $f_+$ and $f_-$  ordered as a function of $f_+ + f_- -f_r$. The red rectangle correspond to data from panel (a). The remaining data are obtained from combined two-tone and single-tone spectroscopy measurements (see Supplementary Material). The green (black) line represents $f_r$ ($f_q$ predicted by the JC model). \textbf{(c)} Discrepancy between the $f+-f-$ experimental (black dots) data and the JC fit, $\delta$. The curves correspond to fits using different theoretical models: the quantum Rabi model (dashed gold), the three-level transmon model (solid green), and the full model given by Eq.~\eqref{eq:full_coupled_hamiltonian} with $U(\hat{\varphi}) = -E_J\cos{\hat{\varphi}}$ (dotted pink). The $E_J/E_C$ ratio is shown for reference and is calculated using the full model.}
\end{figure}

The Hamiltonian of this circuit can be written in terms of the number of Cooper pairs in the island, $\hat{n}$, conjugate to the superconducting phase difference, $\hat{\varphi}$: 
\begin{equation}\label{eq:full_coupled_hamiltonian}
H = 4E_C {\hat{n}}^2 + U(\hat{\varphi}, V_g) + \hbar \omega_r \hat{a}^\dagger \hat{a} + \hbar g_0 \hat{n} (\hat{a}^\dagger + \hat{a}) \;.
\end{equation}
The first two terms represent, respectively, the charging energy of the island, $E_C$, and the Josephson potential $U$, which depends on the gate voltage $V_g$ and, importantly, deviates from that of conventional tunnel junctions used in transmon qubits $U(\hat{\varphi})=-E_J\cos (\hat{\varphi}). $ 
The third term describes the bare resonator which is expressed in terms of its annihilation and creation operators, $\hat{a}$ and $\hat{a}^\dagger$, respectively.
Finally, the capacitive light-matter interaction 
is captured by the last term, where $\hbar g_0 = 2 \beta e V_{\textrm{RMS}}^0$ is a constant that sets the strength of the capacitive interaction and is given by the geometry of the device. Here, $\beta = C_c / C_\Sigma$ is the ratio between the coupling capacitance $C_c$ and the total capacitance to ground $C_\Sigma$ of the gatemon and $V_{\textrm{RMS}}^0$ denotes the resonator's root-mean-square voltage associated with the fundamental mode~\cite{koch_charge-insensitive_2007}. Note that the light–matter coupling, $g$, depends on the qubit transition matrix element via $\hbar g = \hbar g_0 \bra{0}\hat{n}\ket{1}$ and can therefore vary with $V_g$. COMSOL simulations of the superconducting island yield $E_C/h=247$~MHz and a vacuum coupling strength 
$g_0/2\pi=821$~MHz, which would 
place the device in the USC regime. 
The superconducting resonator exhibits a bare resonance frequency $f_r = \omega_r/2\pi = 3.885~\mathrm{GHz}$ and a coupling quality factor to the transmission line $Q_c = 11582$.

\begin{figure*}
\includegraphics[width=170mm]{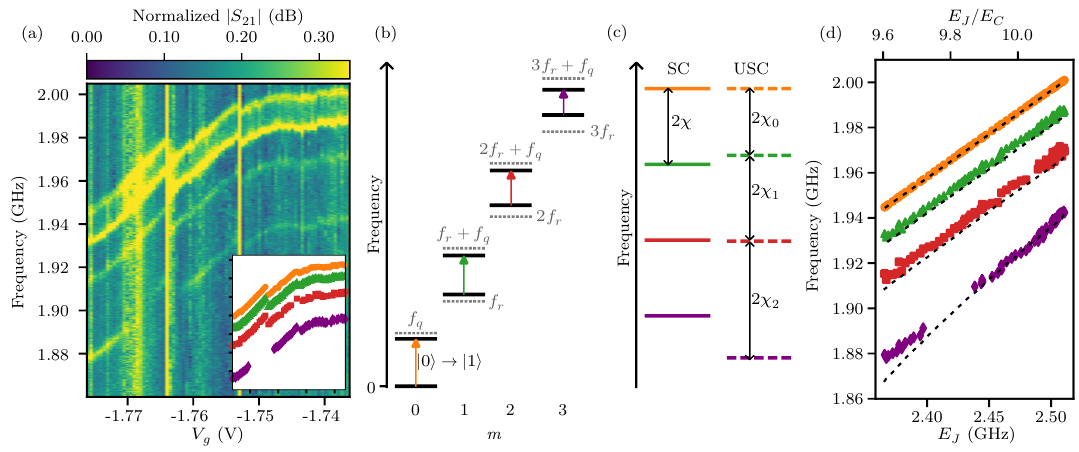}
\caption{\label{fig:coupled_system_spectra} \textbf{(a)} Two-tone spectroscopy as a function of 
$V_g$. The inset shows the extracted peak positions (orange dots, green triangles, red squares and purple diamonds), which heuristically correspond to initial states with $m = 0, 1, 2, 3$ photons in the resonator. \textbf{(b)} Energy-level diagram of the uncoupled (dotted grey) and coupled (solid black) resonator-qubit system for different photon numbers. The colored arrows depict the qubit $|0\rangle \rightarrow |1\rangle$ transition, whose frequency decreases as $m$ increases. \textbf{(c)} Schematic comparison of the dispersive shifts in the strong and ultrastrong coupling regimes. The transition spacing in the latter depends on the resonator photon number. \textbf{(d)} Transition frequencies extracted from (a) compared to the two-tone spectra (dotted-black) calculated using the potential in Eq.~\eqref{eq:abs_potential}.}
\end{figure*}

\textit{Semiconductor Josephson junction.-}
The Josephson potential, \(U(\hat{\varphi},V_g)\), of gatemons is  
defined by high-transmission channels in the semiconductor weak link~\cite{goffman_conduction_2017}, which give rise to Andreev bound states that govern the Josephson transport. 
To account for this, we adopt a minimal effective model based on 
a single-channel short-junction description~\cite{Furusaki1991, Beenakker1992,MartinRodero1994}, with ground state energy given by
\begin{equation}\label{eq:abs_potential}
U(\hat{\varphi}, V_g) = -\Delta_0(V_g)\sqrt{1 - T(V_g)\sin^2(\hat{\varphi}/2)},
\end{equation}
where \(T(V_g)\) is the effective channel transmission and \(\Delta_0(V_g)\) captures microscopic details of the junction, such as the induced gap and the coupling of the normal region to the leads. This model 
captures deviations from the tunnel regime while preserving simplicity, without attempting a full microscopic description. In the transmon regime (\(\expval{\hat{\varphi}} \rightarrow 0\)), a Taylor expansion yields an effective Josephson energy \(E_J(V_g) = \Delta_0(V_g) T(V_g)/4\), and an anharmonicity \(\alpha(V_g) = -E_C(1 - 3T(V_g)/4)\)~\cite{kringhoj_anharmonicity_2018,kringhoj_suppressed_2020,bargerbos_observation_2020,Vakhtel2023}.

\textit{Ultrastrong coupling regime.-}
The coupling strength between the gatemon and the resonator is extracted from the avoided crossing between their energy levels. In our device, this is measured by tuning the Josephson energy, and thus $f_q$, with the gate voltage. 
Figure~\ref{fig:demonstrating_usc}~(a) shows spectroscopy measurements revealing the anticrossing by tracking the absorption peaks associated with the two hybridized modes $f_\pm$. The data underscore the strongly non-monotonic dependence of the hybridized mode frequencies on $V_g$, reflecting the gate-dependence of $E_J$. 
This behavior arises from disorder in the semiconductor and Fabry-Perot oscillations
~\cite{goffman_conduction_2017}.

The JC model provides a convenient starting point for analyzing the data and serves as a reference to quantify deviations from the RWA. 
In the low power limit ($\langle \hat{a}^\dagger \hat{a} \rangle \lesssim 1$), 
the JC model predicts analytical expressions for the two lowest transitions: 
\begin{equation}\label{eq:jc_transitions}
f_{\pm} = \frac{f_r + f_q}{2} \pm \frac{1}{2} \sqrt{{(f_q - f_r)}^2 + {(g/\pi)}^2}.
\end{equation}
From these expressions, $f_q$ can be extracted as: 
\(
    f_q = f_{+} + f_{-} - f_r.
\)
Although this model does not describe the USC regime, we sort our data as a function of this quantity, 
$f_{+} + f_{-} - f_r$. As shown in Fig.~\ref{fig:demonstrating_usc}~(b), this removes the non-monotonicity of the $V_g$-dependence. 
We subsequently fit the avoided crossing using the JC model result,
\(
	f_+ - f_-,
\)
with $g$ as the only fitting parameter. The fit yields $g/2\pi \approx 643$~MHz and 
$g/\omega_r \approx 0.16$
, placing the device in the USC regime, although it does not fully capture the measured dispersion. 

Fig.~\ref{fig:demonstrating_usc}~(c) shows the discrepancy $\delta$ between the experimental data (black circles) and the JC model prediction (dotted gray line). To identify the minimal ingredients required to describe the system, we fit the data using models of increasing complexity
. The dashed yellow curve shows the quantum Rabi model prediction, which includes the counter-rotating terms. The solid-green curve further incorporates the frequency dependence of the coupling, $g$, and the qubit multilevel structure. Finally, the dotted-pink curve shows Eq.~\eqref{eq:full_coupled_hamiltonian} solved numerically, including enough transmon and resonator levels to ensure USC effects are properly captured. Here, we assume a tunnel-junction potential for $U(V_g,\hat{\varphi})$, as the avoided crossing energy scale is not sensitive to the exact shape of the potential (see Supplementary Material). The latter model yields $g_0/2\pi \approx 648$~MHz and $g/\omega_r$ varying between 0.12--0.2 for the $V_g$ range shown in Fig.~\ref{fig:demonstrating_usc}~(c). For details on the numerical fits and how the estimated $g_0/2\pi$ compares to the COMSOL simulations refer to the Supplementary Material. 

\textit{Photon-number-dependent transitions in the USC regime.-}
We now focus on how the USC regime modifies the device spectrum. Fig.~\ref{fig:coupled_system_spectra}~(a) displays two-tone spectroscopy data taken in a $V_g$ range in which multiple adjacent transition lines are observed. The number of visible spectral lines increases with the 
readout power applied to the resonator, as shown in the End Matter. We heuristically attribute these lines to photon-number–dependent dispersive shifts of the qubit $\ket{0}\rightarrow\ket{1}$ transition, $\chi_m \equiv f_{q,m}-f_{q,m+1}$, arising from different resonator photon populations $m$ (Fig.~\ref{fig:coupled_system_spectra}~(b))~\cite{Schuster2007}, although we emphasize that for USC, $m$ is no longer a good quantum number due to strong hybridization.

The dispersive approximation in the SC regime predicts $m>0$ transitions to shift by $2m\chi$ with respect to the $m=0$ case. 
The USC however effectively prevents the system from entering the dispersive regime and higher order terms need to be included, adding a dependency on $m$. This results in the set of unequally spaced $\chi_m$ transitions observed in the measurement, as schematically depicted in Fig.~\ref{fig:coupled_system_spectra}~(c).

\begin{figure}
\includegraphics[width=86mm]{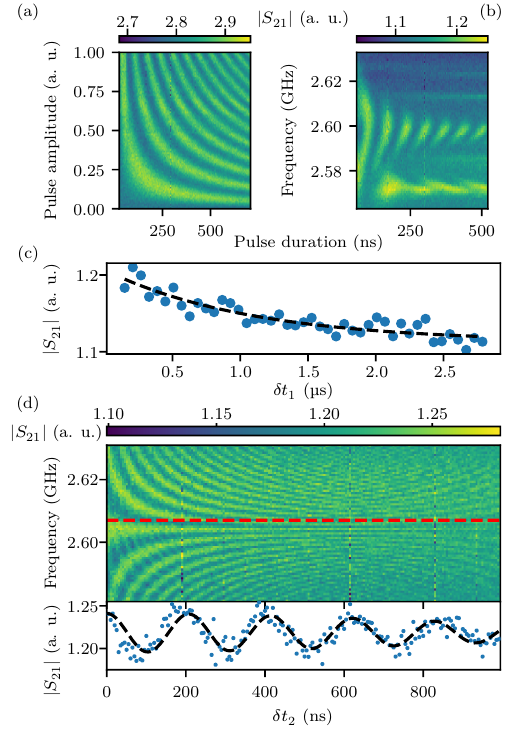}
\caption{\label{fig:time_resolved} Coherent Rabi oscillations as a function of \textbf{(a)} the drive pulse duration and amplitude ($f_q = 2.598$~GHz and $V_g=3.184$~V), and \textbf{(b)} the drive frequency and pulse duration ($V_g = 3.174$~V). \textbf{(c)} $T_1$ relaxation measurement taken at same $V_g$ and $f_q$ as (a). \textbf{(d)} Ramsey oscillations as a function of drive frequency and delay between two consecutive $\pi/2$ pulses. Lower panel displays a line cut at a drive frequency of 2.607 GHz. Measurement taken at the same $V_g$ as (b). All measurements were taken around 
the same 
sweet spot with small variations in $V_g$ and $f_q$ arising from gate hysteresis and frequency drift.}
\end{figure}

For a more quantitative analysis, the expected spectra are calculated using the full Hamiltonian in Eq.~\eqref{eq:full_coupled_hamiltonian}, taking the previously fitted value of $g_0$ and the potential given in Eq.~\eqref{eq:abs_potential}. The result is plotted in Fig.~\ref{fig:coupled_system_spectra}~(d) (dotted lines), together with the experimental data (markers). To perform the fit, $T$ is treated as a free parameter, yielding good agreement with the data for $T=1$, which results in $\alpha=-E_C/4$. Regarding the parameters $E_J$, $\alpha$ and $T$ entering the model, we note that the junction is long compared to the Al-shell coherence length of the order of 100~nm. In this regime, finite-length effects are expected to suppress both \(E_J\) and \(\alpha\) with respect to the short-junction limit~\cite{Fatemi25, Bagwell1992, Tosi19}. Consequently, these parameters should be regarded as effective and are not directly related to microscopic quantities such as the superconducting gap. 

\textit{Qubit dynamics in the USC regime.-}
Finally, we demonstrate coherent time-domain control of the device in the USC regime and investigate its decay and decoherence properties. 
In Fig.~\ref{fig:time_resolved}~(a), Rabi oscillations are measured by driving the qubit at frequency \( f_q \) 
as a function of pulse duration and amplitude. Fig.~\ref{fig:time_resolved}~(b) shows the characteristic Chevron pattern arising from varying drive frequency near the qubit resonance. Additional oscillations can be observed at lower frequencies corresponding to the qubit transition occuring while having a photon number $m=1$ in the resonator. Fig.~\ref{fig:time_resolved}~(c) shows a relaxation-time (\(T_1\)) measurement, where a \(\pi\) pulse excites the qubit and its response is recorded as a function of the waiting time \(\delta t_1\) as it relaxes to the ground state. A fit to the data yields \(T_1 = 1.1~\si{\micro\second}\). Finally, Fig.~\ref{fig:time_resolved}~(d) presents a Ramsey interferometry experiment used to extract the decoherence time \(T_2^*\). Two \(\pi/2\) pulses separated by a delay \(\delta t_2\) are applied at different drive frequencies. Fitting the exponentially decaying Ramsey oscillations with a Gaussian envelope yields \(T_2^* = 1.2~\si{\micro\second}\).

Compared to other gatemon qubits based on InAs/Al nanowires, our $T_1$ values are consistent with early demonstrations~\cite{larsen_semiconductor-nanowire-based_2015, danilenko_few-mode_2023, feldstein-bofill_gatemon_2024, sabonis_destructive_2020}, though they remain below the best reported implementations, which have reached up to 5~\si{\micro\second}~\cite{casparis_gatemon_2016} and 15~\si{\micro\second}~\cite{luthi_evolution_2018}. Regarding the decoherence time $T_2^*$, our results fall short of the highest reported values—up to 4~\si{\micro\second}~\cite{casparis_gatemon_2016, luthi_evolution_2018}—but are comparable to those observed in InAs/Al gatemons reported in the literature~\cite{larsen_semiconductor-nanowire-based_2015, casparis_gatemon_2016}. To extend the picture provided by the measurements in Fig.~\ref{fig:time_resolved}, we further characterize $T_1$ and $T_2^*$ as a function of $V_g$ at a different sweet spot, revealing performance of the same order of magnitude as shown here (see End Matter).

\textit{Conclusions.- } We have demonstrated a gatemon qubit in the USC regime using capacitive coupling within a standard coplanar geometry. 
Spectroscopic measurements reveal photon-number–dependent transitions that deviate from the JC ladder expected in the strong coupling regime, providing clear signatures of ultrastrong light–matter interaction. In addition, we demonstrate time-resolved dynamics with $T_1$ and $T_2^*$ comparable to those previously reported in the literature, suggesting that relaxation and decoherence are dominated by the material platform~\cite{feldstein-bofill_gatemon_2024} rather than by Purcell losses via the resonator~\cite{purcell_resonance_1946}, which are enhanced in the USC regime. These results encourage further exploration of time-resolved dynamics in the USC regime and highlight the potential of hybrid semiconductor–superconductor circuits for exploring new device concepts~\cite{pita-vidal_novel_2025}.

\begin{acknowledgments}
    \textit{Acknowledgments.- } We acknowledge funding from the EU through the European Research Council (ERC) Starting Grant agreement 716559 (TOPOQDot), through the European Innovation Council Pathfinder grant no. 101115315 (QuKiT), and through the "NextGenerationEU"/PRTR and "ERDF A way of making Europe" programs. We also acknowledge funding from the Danish National Research Foundation (DNRF 101), the SolidQ project of the Novo Nordisk Foundation, the Carlsberg Foundation, from the Spanish MCIN/AEI/10.13039/501100011033 institutions through Grants No.~PID2023-150224NB-I00, PID2024-161156NB-I00, PID2022-137779OB-C41, PID2022-137779OB-C42 and FPU20/01871, the ``Mar\'{\i}a de Maeztu'' Program for Units of Excellence in R\&D (CEX2018-000805-M), and the ``Severo Ochoa'' Centers of Excellence program through Grants CEX2024-001445-S and CEX2020-001039-S, and from the Comunidad de Madrid through Grant PIPF-2022/TEC-24452.

\end{acknowledgments}

\bibliography{main.bib}

\appendix

\onecolumngrid
\begin{center}
    \vspace{3ex}
    \textbf{End Matter}
\end{center}
\twocolumngrid

\begin{figure}
\includegraphics[width=86mm]{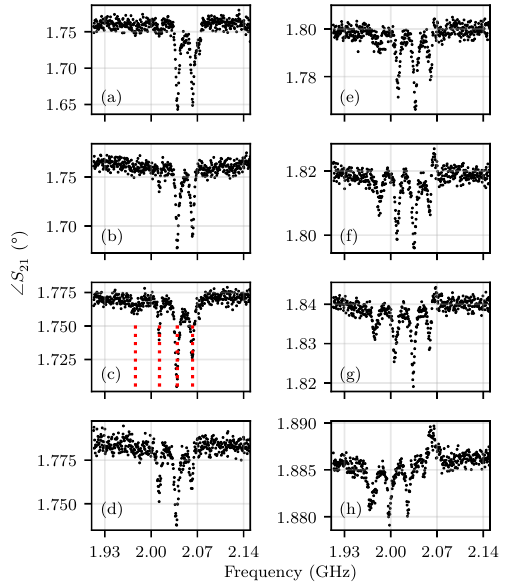}
\caption{\label{fig:photon_number_cuts} Phase of the resonator $S_{21}$ signal measured while applying a tone at $f_r - 2~\si{\mega\hertz}$ and as a function of the frequency of a drive tone applied near the qubit transition frequency. Panels (a)-(h) correspond to increasing input power of the first tone in 3 dB steps. Dotted red lines in (c) indicate the expected position of the photon-number–resolved transitions.}
\end{figure}
\textit{Photon-number-dependent qubit transitions.- }
To identify the origin of the transitions in Fig.~\ref{fig:coupled_system_spectra}~(a), we perform additional measurements for a similar $f_q$, as shown in Fig.~\ref{fig:photon_number_cuts}. The measurements reveal that increasing the input power at a frequency near the resonance of the superconducting cavity, 
leads to the emergence of additional absorption peaks. Following Ref.~\cite{Schuster2007}, we interpret these peaks as transitions related to higher resonator photon occupancies. Thus, they are a result of photon-number-dependent shifts, $\chi_m$, of the qubit transition, and are resolved because $\chi_m$ exceeds the linewidths of both the qubit and the resonator.

At low input power, the expected position for the 
photon-number-dependent transitions can also be obtained from the same full Hamiltonian model described in the main text, Eq.~\eqref{eq:full_coupled_hamiltonian}, including the ABS potential. In particular, for panel (c), a good agreement with the data is obtained for $T = 0.95$. However, panels (e)-(h) reveal that the spacing between peaks does not remain constant with the input power applied to the resonator. In this case, 
a full quantum treatment of the system dynamics needs to be considered to fully describe the observed spectra~\cite{tosi_effects_2024}.

\begin{figure}
\includegraphics[width=86mm]{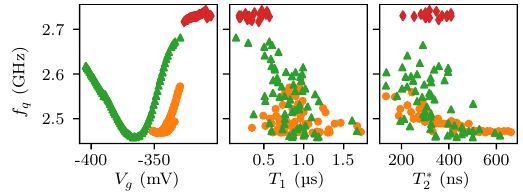}
\caption{\label{fig:t1_t2_times} \textbf{(a)} Points at which $T_1$ and $T_2^*$ values have been obtained with different $f_q(V_g)$ sweeps. \textbf{(b)} and \textbf{(c)} display the values of $T_1$ and $T_2^*$, respectively, measured in the $f_q(V_g)$ points indicated by (a). Different colors and shapes indicate distinct $V_g$ sweeps, voltage hysteresis effects can be observed between the green and orange datasets.}
\end{figure}

\textit{Gate-dependence of relaxation and decoherence times.- }
Relaxation and coherence times are influenced by several mechanisms, including contributions that depend on the applied $V_g$. For example, Purcell decay depends on the detuning between the qubit and the resonator, which can be tuned electrostatically. Moreover, due to the semiconducting nature of the weak link, $E_J$ is sensitive to the electrostatic environment, and thus to charge noise around the nanowire junction. To gain further insight into the qubit in the USC regime, we measure relaxation and decoherence times as a function of $V_g$.  We focus on a minimum of $f_q(V_g)$, shown in Fig.~\ref{fig:t1_t2_times}~(a), which defines a sweet spot where first-order sensitivity to gate voltage fluctuations vanishes~\cite{vion_manipulating_2002,casparis_gatemon_2016}. $T_1$ ranges between 0.5--1~\si{\micro\second} across the measured gate voltages, with a modest enhancement near the sweet spot [Fig.~\ref{fig:t1_t2_times}~(b)]. In contrast, $T_2^*$ shows a more pronounced improvement, reaching 400--650~\si{\nano\second} near the sweet spot, compared to 200--400~\si{\nano\second} away from it [Fig.~\ref{fig:t1_t2_times}~(c)], consistent with reduced sensitivity to low frequency electrostatic noise~\cite{ithier_decoherence_2005,krantz_quantum_2019}. Notably, the measurements in the main text were taken at a different local minimum of $f_q(V_g)$, less detuned from the resonator, yet exhibit larger $T_2^*$ values than those in Fig.~\ref{fig:t1_t2_times}~(c). Taken together, these observations indicate that electrostatic noise locally affecting the semiconductor Josephson junction is the dominant decoherence channel, with the USC regime playing a secondary role in setting the observed timescales of order 1~\si{\micro\second}.

Therefore, the timescales reported here seem to be primarily limited by the material platform rather than by the USC regime itself. Improving coherence in future devices will require e.g., improving materials, and optimizing design and fabrication to reduce the coupling of the nanowire junction to electrical noise. 
Understanding and mitigating decoherence in hybrid semiconductor-superconductor devices remains an open challenge, with recent work exploring strategies to suppress low-frequency noise and optimize material combinations for more robust qubits~\cite{feldstein-bofill_gatemon_2024, purkayastha_transmon_2025}. Our results establish gate-tunable USC as a viable platform for qubit operation, and point to electrostatic noise engineering as the key lever for improving coherence in this regime.

\end{document}